\documentclass[10pt]{iopart}
\usepackage{multirow}
\usepackage{graphicx}
\usepackage{amsmath}

\begin{document}

\title[Measuring time-resolved heat transfer fluctuations on a heated-thin foil in a turbulent channel airflow]{Measuring time-resolved heat transfer fluctuations on a heated-thin foil in a turbulent channel airflow}

\author{A Cuéllar$^1$, E Amico$^2$, J Serpieri$^2$, G Cafiero$^2$, W J Baars$^3$, S Discetti$^1$, A Ianiro$^1$}

\address{$^1$ Department of Aerospace Engineering, Universidad Carlos III de Madrid, Leganés, Spain}
\address{$^2$ Department of Mechanical and Aerospace Engineering, Politecnico di Torino, Torino, Italy}
\address{$^3$ Faculty of Aerospace Engineering, Delft University of Technology, Delft, The Netherlands}

\ead{acuellar@ing.uc3m.es}

\vspace{10pt}
\begin{indented}
\item[]Received 16 October 2024, revised 7 February 2025, accepted for publication 28 February 2025
\item[]\textbf{Cite this work as:} A Cu\'ellar, E Amico, J Serpieri, G Cafiero, W J Baars, S Discetti and A Ianiro 2025 Measuring time-resolved heat transfer fluctuations on a heated-thin foil in a turbulent channel airflow \emph{Meas. Sci. Technol.} \textbf{36}(4) 045303
\item[]\textbf{DOI:} \texttt{10.1088/1361-6501/adbb84} (https://doi.org/10.1088/1361-6501/adbb84)
\end{indented}

\begin{abstract}
We present an experimental setup to perform time-resolved convective heat transfer measurements in a turbulent channel flow with air as the working fluid. We employ a heated thin foil coupled with high-speed infrared thermography. 
The measurement technique is challenged by the thermal inertia of the foil, the high frequency of turbulent fluctuations, and the measurement noise of the infrared camera. We discuss in detail the advantages and drawbacks of all the design choices that were made, thereby providing a successful implementation strategy to obtain high-quality data.
This experimental approach could be valuable for studies employing wall-based measurements of turbulence, such as flow control applications in wall-bounded turbulence.

\end{abstract}

%

\vspace{1pc}
\noindent\fbox{\parbox{\dimexpr\textwidth-2\fboxsep-2\fboxrule}{\small
\textbf{Accepted Manuscript.} This is the Accepted Manuscript version of an article accepted for publication in \emph{Measurement Science and Technology}. IOP Publishing Ltd is not responsible for any errors or omissions in this version of the manuscript or any version derived from it. The Version of Record is available online at https://doi.org/10.1088/1361-6501/adbb84.

\vspace{0.5pc}
\noindent\textbf{Copyright and licence.} \copyright\ 2025 IOP Publishing Ltd. This Accepted Manuscript is available for reuse under a CC BY-NC-ND 4.0 licence (https://creativecommons.org/licenses/by-nc-nd/4.0/) after the 12-month embargo period, provided that all the terms of the licence are adhered to. The embargo period ended on 20 March 2026.}}

\maketitle

\ioptwocol

\section{Introduction}


In this work, we present an experimental setup for conducting time-resolved measurements of the convective heat transfer coefficient on the wall of a turbulent channel. Measurements are based on the unsteady heated-thin foil sensor \cite{nakamura2009frequency, nakamura2013quantitative}, coupled with infrared (IR) thermography as a temperature transducer. In addition, to do this in a non-intrusive way, the heating applied must be small enough to avoid perturbing the flow with undesired buoyancy forcing. Excessive heating could increase the Richardson number (ratio of natural to forced convection), potentially hindering the results. This paper responds to a challenge of measurements in air flows. Due to the foil thermal inertia and the characteristic frequencies of the problem, the measurement of the small amplitude temperature fluctuations requires careful image pre-processing, including spatial, temporal and feature-based filtering. We critically discuss all the design choices made, together with the details of image processing to obtain high-quality instantaneous measurements of the Stanton number in air with this technique. 

IR thermography has been applied in different experimental setups to study heat transfer by a wall-bounded grazing flow. Following the seminal work by Hetsroni and Rozenblit \cite{hetsroni1994heat}, Gurka et al. \cite{gurka2004detecting} conducted synchronised measurements of particle image velocimetry (PIV) and heat transfer over a hot foil subject to a grazing turbulent wall-bounded water flow. A heated thin foil was employed on a similar boundary layer problem in water with high-frequency IR measurements in synchronization with PIV \cite{foroozan2023synchronized}. Similar experiments with air as the working fluid are challenged by lower fluctuations of the foil temperature and heat transfer, and higher frequency content of said fluctuations. Both of these aspects require a push towards thermally thinner and tailored processing of the images. The first challenge is merely technological. On the other hand, image conditioning requires careful consideration of the choice of the processing strategy, considering that the signal-to-noise ratio can easily be below 1.

\begin{figure}[ht]
\centering 
    \includegraphics[trim= 0 5cm 0 0, width=\linewidth]{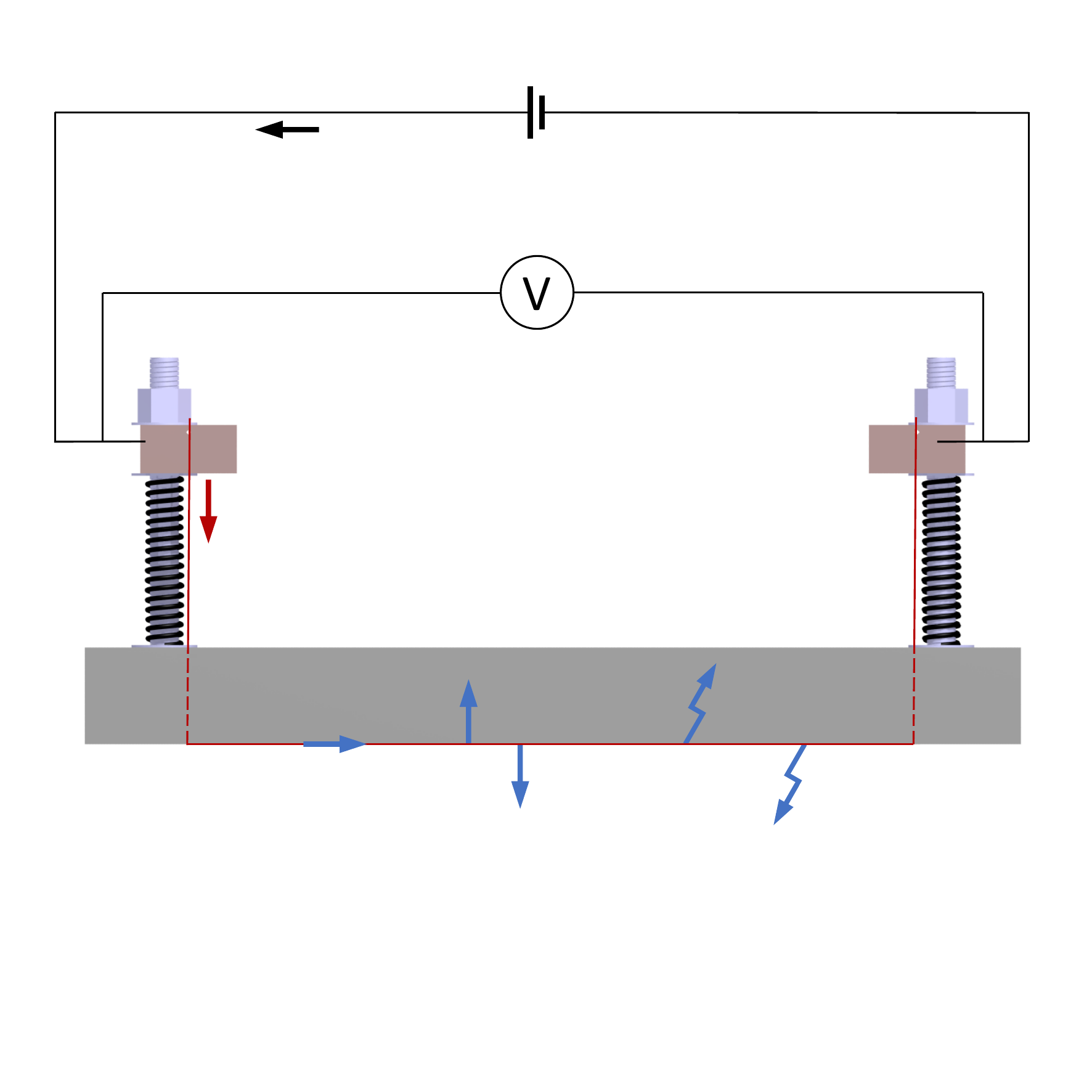}
    \put(-165,160) {$I$}%
    \put(-190,70) {$\phi^{''}_{\rm J}$}%
    \put(-175,20) {$\phi^{''}_{\rm cond}$}%
    \put(-155,60) {$\phi^{''}_{\rm conv,ext}$}%
    \put(-140,8) {$h_c(T_{\rm w}-T_{\rm aw})$}%
    \put(-75,45) {$\phi^{''}_{\rm rad}$}%
\caption{Sketch of the electric circuit mounted on the sensor and the heat fluxes on the energy balance of the thin foil.}
\label{fig:energybalance}
\end{figure}
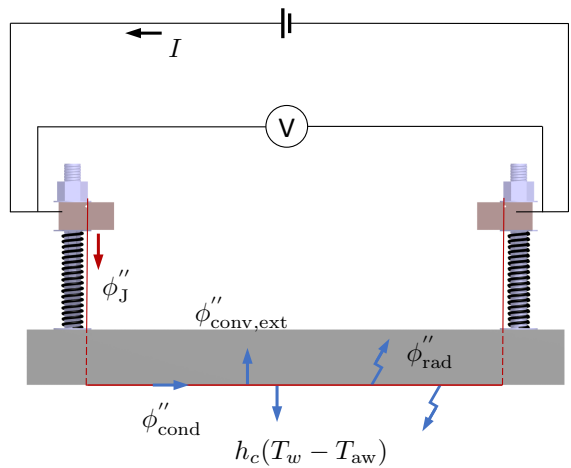

The heated-thin foil sensor is based on measuring the convection between the fluid and a heated foil through an energy balance (equation \ref{eqn:energybalance}), as sketched in figure \ref{fig:energybalance}. It assesses the temporal temperature variation of the foil as an unsteady term with the contribution of the different instantaneous heat fluxes and sources, where $c_p$ is the specific heat capacity, $\rho$ is the density and $a$ is the thickness.

\begin{equation}
    c_p \rho a  \frac{\partial T_{\rm w}}{\partial t} = \phi^{''}_{\rm J} - \phi^{''}_{\rm cond} - 2\phi^{''}_{\rm rad} - \phi^{''}_{\rm conv,ext} - h_c(T_{\rm w}-T_{\rm aw}) ~. \label{eqn:energybalance}
\end{equation}
The superscripts ($''$) denote that heat fluxes are expressed per unit area of the thin foil. These terms include the heat input provided to the sensor (in most cases produced by the Joule effect) $\phi^{''}_{\rm J}$, the conduction within the foil $\phi^{''}_{\rm cond}$, the radiation $\phi^{''}_{\rm rad}$ emitted from both sides of the foil, and the convection, to be treated separately for the side exposed to the flow, equal to $h_c(T_{\rm w}-T_{\rm aw})$ and for the side not exposed to the flow $\phi^{''}_{\rm conv, ext}$, where natural convection may develop. The convective heat flux between the flow and the wall depends on the convection heat transfer coefficient $h_c$, which is the quantity to be determined, and on the local wall temperature $T_{\rm w}$ and the adiabatic wall temperature $T_{\rm aw}$, which are both measured with the IR camera. Terms on the right-hand-side of equation \ref{eqn:energybalance} are typically modelled following conduction and radiation laws, along with empirical correlations (see, e.g. those reported in \cite{bergman2020fundamentals}).

The thermal model relies on the assumption that the foil is thermally thin, such that the temperature on the internal side of the foil (considered in the model) is equal to that on the external side (measured temperature). This approximation is valid if the Biot number, $Bi = \tfrac{h_c a}{\kappa}\ll 1$ (where $\kappa$ is the foil thermal conductivity). Additionally, the Fourier number $Fo = \tfrac{\alpha t_{{\rm char}}}{a^2}$ (with $t_{{\rm char}}$ being the characteristic time of the problem) compares the heat flux and the rate of thermal energy storage, requiring $Fo\gg1$ to perform unsteady heat transfer measurements \cite{astarita2012infrared}. Both requirements are met in the experimental setup presented in the current work, owing to a small enough foil thickness, $a$.

The current experimental approach presents a series of challenges that must be carefully managed to ensure accurate results.
Provided that the available IR camera is capable of an acquisition frequency large enough to temporally resolve the flow scales involved in the problem, several difficulties are still to be faced.
First of all, the IR camera must be sensitive enough to detect the wall temperature fluctuations.
While most of the terms are mainly steady, the temporal variations of $h_c$ lead to temperature fluctuations of the foil, which are damped by the foil thermal inertia on the left-hand side of equation (\ref{eqn:energybalance}). Secondly, the fluctuations to be measured must be compared to the Noise Equivalent Temperature Difference (NETD) of the camera to evaluate the IR sensor suitability \cite{greco2014time}. The characteristic time\textemdash or the characteristic frequency\textemdash of the problem is a property of the flow and can be estimated. For instance, from dimensional analysis, it can be quantified as the ratio of the characteristic length and the characteristic velocity. As higher frequencies make the unsteady term cause smaller temperature fluctuations, given such characteristic time, one may want to increase the temperature fluctuations in case they are too small to be detected. Therefore, from equation \ref{eqn:energybalance} it is possible to argue that one can either increase the foil heating, which in turn leads to greater natural convection, or choose a foil thin enough to amplify the temperature fluctuations. The thickness of the foil is a critical factor, however, it cannot be chosen \textit{ad libitum} as the foil needs to be manufacturable, robust enough to be implemented and sustain pressure and shear fluctuations without deforming, and ideally be cost-effective.
Moreover, for IR measurements, a layer of paint is often applied to enhance the accuracy of temperature measurements by increasing the surface emissivity. This alters the thermal properties of the material and its thickness and thus may distort the measurements if not properly accounted for \cite{stafford2009characterizing}. A proper balance between these factors is essential for reliable outcomes.

Different models have been developed to address these complexities \cite{nakamura2009frequency, stafford2009characterizing, torre2018evaluation, astarita2012infrared}. However, for measurements in air flows, the temperature fluctuations are often of the same order (or smaller) than the NETD, thus making image preprocessing crucial. Techniques like Proper Orthogonal Decomposition (POD) or autoencoders can suppress noise in very low signal-to-noise ratio conditions \cite{raiola2017towards, gu2024denoising}. Additional techniques, such as bad pixel removal to eliminate outliers and detrending procedures to counteract increasing mean temperatures in the system, further improve data quality \cite{foroozan2023synchronized}. 

The remainder of this article contains a description of the experimental setup in section \ref{sec:experiment}, a discussion of the results in section \ref{sec:results}, the uncertainty is quantified in section \ref{sec:uncertainty} and the key findings are summarized in section \ref{sec:conclusions}.

\section{Experimental setup} \label{sec:experiment}

\subsection{Channel flow and experimental setup} \label{sec_channel_flow}

\begin{figure}[ht]
\centering
\includegraphics[trim = 1cm 0 3cm 0, width=\linewidth]{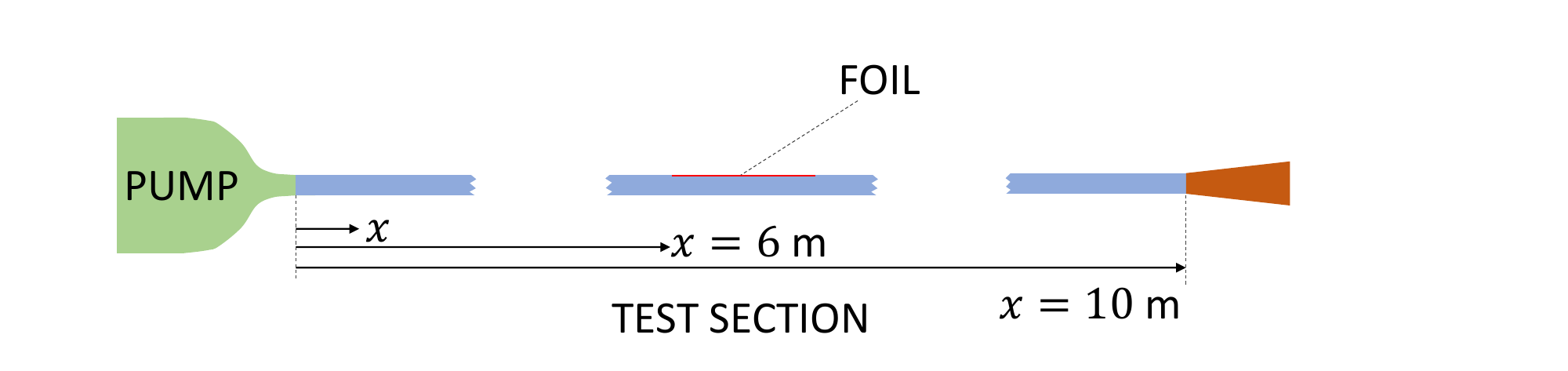}
\caption{Sketch of the channel flow facility, with the thin foil (red) mounted on the top wall of the channel.}
\label{fig:channelsketch}
\end{figure}

The experiments of this work were performed in the channel flow facility of the aerodynamics laboratory at Politecnico di Torino. The walls of the channel are made of poly-methyl-methacrylate, 
defining an internal cross-section of 35 (height) $\times$ 420 (width) mm$^2$ with a length of 10 m. The channel operates with air in an open-return configuration. The inlet section is composed of an electric pump followed by a small divergent section, a settling chamber and a convergent nozzle upstream of the channel entrance. To trigger transition to turbulence, two thin turbulator strips with zig-zag-shaped leading edges are placed at the end of the convergent on the channel's top and bottom walls, along the spanwise direction. To ensure flow stability, a divergent diffuser is installed at the outlet of the channel. The channel walls are made of modular sections which allow a custom layout to install suitable devices needed to carry out the experimental campaigns. A schematic representation of the channel is reported in figure \ref{fig:channelsketch}.

The experiments reported in this work have been conducted in a turbulent regime with bulk airspeed $U_\infty= 4.95$ m/s. The Reynolds number based on the outer quantities is equal to $Re=\frac{U_\infty h}{\nu}=5800$, where $h$ is the half channel height, equal to 17.5 mm and $\nu$ the air kinematic viscosity. Accordingly, the friction-based Reynolds number $Re_{\tau}=\frac{u_\tau h}{\nu}=220$, with $u_\tau=\sqrt{\frac{\tau_w}{\rho}}$ the friction velocity defined as the square root of the ratio between the wall skin friction $\tau_w$ and the fluid density $\rho$. The friction velocity is characterised by employing 16 static pressure ports distributed along the channel test section to quantify the pressure gradient. The static pressure measurements are carried out using a 16-channel DSA pressure scanner, with a maximum pressure range of 2500 Pa and a 0.05\% full-scale accuracy. A linear fitting is applied to obtain the pressure gradient, which is then employed to determine the value of the wall friction \cite{serpieri2024conditioning}.

\subsection{Heat transfer measurements} \label{sec.heattransfer}

\begin{figure}[ht]
\centering
    \includegraphics[trim = 0 0.8cm 0 0.8cm,width=\linewidth]{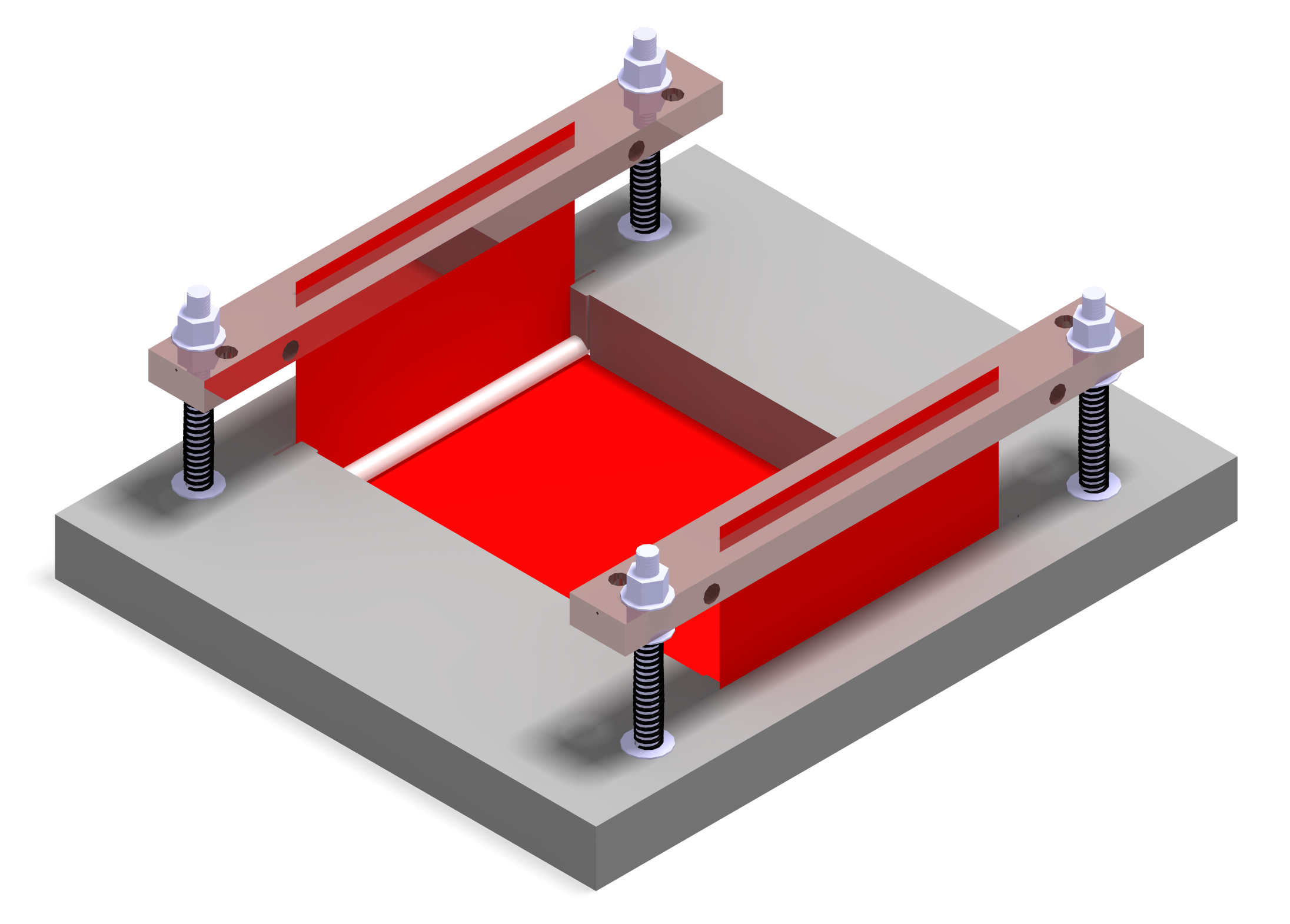}
\caption{Isometric view of the external side of the sensor, with the thin foil (red) mounted on the frame.}
\label{fig:sensor}
\end{figure}

A heated-thin-foil heat transfer sensor has been designed and mounted on the upper wall of the channel (see figure \ref{fig:sensor}) by means of a frame embedded in the top wall, coincident with the position of the modular wall it replaces. This design choice minimizes the flow disturbance caused by buoyancy effects. Indeed, the intensity of natural convection for a heated plate facing downward is lower than when facing upward \cite{wong1977handbook}. 

The sensor frame was 3D printed with PLA material. 
The thin foil is inserted on the frame through two groves at its leading and trailing edge positions, 150 mm apart. The sensing area has a width $W=100$ mm and it spans a length $L=150$ mm along the channel wall.
On the external part of the frame, four fixed bars are mounted, holding two copper block pairs, one for each end of the thin foil. Each pair of copper blocks forms a clamp that retains the thin foil in between. 
The thin foil is heated through the Joule effect. The uniform thickness of the foil ensures constant heat flux; temperature measurements of the heated foil surface confirmed that the temperature distribution was nearly spatially uniform. To apply Joule heating, a DC power supply is connected to the copper blocks. The higher electrical resistance of the CrNi-Steel alloy with respect to that of the copper, together with the small thickness of the thin foil (5 $\mu$m) with respect to that of the copper blocks (1 cm) ensures that the electric potential drop through the copper is negligible with respect to that through the thin foil. The copper blocks can thus be considered at practically constant voltage, ensuring uniform voltage and current in the spanwise direction. To minimize contact resistance between the copper blocks and the foil, a thin engraving in the copper block face in contact with the foil is filled with a 1 mm indium wire. The voltage differential applied to the foil is thus assumed to be equal to that between the copper blocks at the foil edges and it is measured with a voltmeter in contact with the copper blocks. The different parameters of the problem employed for this model are collected in table \ref{tab:foilparams}. Voltage and current are not included as several power levels are considered in this work.

\begin{table}[] 
\centering
\caption{\label{tab:foilparams}Values used in the thermal model of the sensor. Values and tolerances of these material properties have been determined from manufacturers' technical information.}
\begin{indented}
\item[]\begin{tabular}{llll}
\br
Quantity & Symbol & Value & Units \\ \mr
Heat capacity (foil) & $c_{p,f}$ & 500 $\pm$ 5 & J/(kg K)\\
Heat capacity (paint) & $c_{p,p}$ & 5000 $\pm$ 50 & J/(kg K)\\
Density (foil) & $\rho_f$ & 7900 $\pm$ 50 & kg/m$^3$ \\
Density (paint) & $\rho_p$ & 1300 $\pm$ 50 & kg/m$^3$ \\
Therm. cond. (foil) &  $\kappa_f$  & 17 $\pm$ 2 & W/(m K) \\
Therm. cond. (paint) &  $\kappa_p$  & 1.4 $\pm$ 0.1 & W/(m K) \\
Area (Joule) & $A^*$ & 0.0254 $\pm$ 0.001 & m$^2$  \\
Foil length & $L$ & 0.15 $\pm$ 0.0005 & m  \\
Foil width & $W$ & 0.1 $\pm$ 0.0005 & m  \\
Emissivity (paint) & $\epsilon$ & 0.95 $\pm$ 0.02 & -- \\
Ambient temperature & $T_{\rm amb}$ & 19 & ºC \\
\br
\end{tabular}
\end{indented}
\end{table} 

To ensure foil tension, a spring is mounted around each bar, pushing apart the sensor frame and the copper blocks, thus tensioning the thin foil. 
The frame adjusts the thin foil, subjecting it to tensile stresses to keep it flat and prevent any misalignment with the rest of the channel wall. The thin foil is made of 1.4310 CrNi-Steel alloy (Cr 16--18\% and Ni 7--9\%) able to withstand a nominal tensile stress $F>1500$ N/mm$^2$. 
Kapton$^\text{\textregistered}$ tape is used for sealing between the sensor foil and the frame, on the flow exposed side. This ensures a hydrodynamically smooth surface and avoids any air leakages perturbing the flow. To assess a suitable foil thickness two foils are tested, with $a_f=5$ $\mu$m and 10 $\mu$m, respectively.

An IR camera is employed as a temperature transducer to measure the temperature on the external side of this sensor. To enhance IR temperature measurements, given the low emissivity of steel, the external side of the thin foil is sprayed with high-emissivity matt black paint.

The IR camera used in this experiment as temperature detector is an Infratec Camera ImageIR$^\text{\textregistered}$ 6300Z, with a resolution of 640 $\times$ 512 pixel$^2$. The temperature resolution (NETD) of the IR camera at $30^{\circ}$C is 0.03 K, and its calibration accuracy is $\pm$2 K. It is worth noting that, following equation (\ref{eqn:unsteady}), the present measurement technique makes use of temperature differences rather than of absolute temperatures, thus the typical uncertainty to be taken into account is the temperature resolution. The camera is mounted at a distance of 30 cm from the foil, with a focal length of 18 mm. This leads to a resolution of 0.21 mm/pixel along the sensor. An integration time equal to 2900 $\mu$s was set according to the configuration requirements of the camera hardware for the range of temperatures considered for this experiment. The IR camera sampling frequency was set to $f_s=180$ Hz to ensure sufficient temporal resolution. This leads to a temporal separation between snapshots of less than 1/16$^{\rm th}$ of an eddy turnover time $h/u_\tau$. Considering a convection velocity of near-wall streaks of about $11u_\tau$ \cite{del2009estimation} and the characteristic length of the sensor $L$, the sampling rate is sufficiently large so that 12 snapshots cover the convective time of the flow over the streamwise length of the sensor. 
%


As a standard procedure of the heated-thin-foil sensor to obtain the time-averaged mean adiabatic wall temperature $T_{\rm aw}(x,z)$, an acquisition run without power input is carried out. Then, the electrical input is turned on to acquire the time-resolved temperature maps of the foil, $T_{\rm w}(x,z,t)$. 

The different heat fluxes in the thin foil are modelled to compute the heat transfer coefficient between the foil and the channel flow from the energy balance equation (\ref{eqn:energybalance}). The unsteady and conductive terms have been corrected to account for the effect of the paint \cite{raghu2006thermal, susa2008influence}, as shown in equations (\ref{eqn:unsteady}) and (\ref{eqn:conduction}) respectively, assuming a coating thickness $a_p=20$ $\mu$m on each side of the foil and using the foil ($f$) and paint ($p$) properties:
\begin{equation}
    c_p \rho a \frac{\partial T_{\rm w}}{dt} = \left( c_{p,f} \rho_f a_f + 2 c_{p,p} \rho_p a_p \right) \frac{\partial T_{\rm w}}{dt} ~. \label{eqn:unsteady}
\end{equation}

The Joule effect term is given by:
\begin{equation}
    \phi^{''}_J = \frac{VI}{A^*}~,
\end{equation}
where $V$ and $I$, respectively, are the voltage and the intensity of the current applied, and $A^*$ is the area through which the current is discharged. Note that this area does not coincide with the flow exposed area of the sensor, as the entire foil surface area between the two pairs of copper blocks must be considered.

The conduction heat flux experienced through the foil is computed as:
\begin{equation}
    \phi^{''}_{\rm cond} = - \left( \kappa_f a_f + \kappa_p a_p \right) \nabla^2T_{\rm w}~, \label{eqn:conduction}
\end{equation}
where $\kappa_f$ and $\kappa_a$ are the thermal conductivities. The nabla operator $\nabla$ is used to indicate the second derivatives of the temperature maps along both principal directions $(x,z)$ of the plane of the foil.

For the quantification of the radiation effect, the emissivity $\epsilon$ of the surfaces of the thin foil should be introduced in the model. As both sides are covered with matt black paint, a high value (0.95) is taken\textemdash the external side is painted to improve its emissivity for IR acquisition purposes, and the internal side is painted to avoid reflections on synchronized PIV acquisition, which is not pertinent to this work. Applying Stefan-Boltzmann's law, radiation is modelled as:
\begin{equation}
    \phi^{''}_{\rm rad} = \sigma \epsilon (T_{\rm w}^4-T_{\rm amb}^4)~, \label{eqn:radiation}
\end{equation}
being $\sigma$ the Stefan-Boltzmann constant and $T_{\rm amb}$ the ambient temperature in the laboratory during the experimental acquisition used as a reference for radiation.

\subsection{Image preprocessing} \label{sec:datafiltering}

Due to the low signal-to-noise ratio, in combination with the need to compute 2$^{\rm nd}$ order derivatives (equation \ref{eqn:conduction}) and fourth-order terms of temperature (equation \ref{eqn:radiation}), preprocessing of the images is of paramount importance. The following steps are implemented:
\begin{itemize}
    \item The effect of the heating of the foil frame and of the natural convection cells rising from the external side of the thin foil was visible on the sequence of original images, characterised by large patterns (characteristic length-scale of the order of the width of the thin foil) and low-frequency temperature oscillations. Due to the low-frequency nature of natural convection, a high-pass filter can be applied, removing it from the fields. To avoid filtering any actual feature coming from the turbulent flow, features whose characteristic time is at least 12 times larger than one eddy turnover time ($h/u_\tau$) have been removed, thus, with a cutoff frequency of 0.9 Hz for this case. 
    Consequently, the term $\phi^{''}_{\rm conv,ext}$ in equation (\ref{eqn:energybalance}) should be neglected for the calculation of $h_c$. This effect can be quantified according to a horizontal heated plate facing up.
    The filter cannot suppress characteristic patterns of the channel turbulence, as with a convection velocity of near-wall streaks of about $11u_\tau$ \cite{del2009estimation}, the filter would suppress structures with a characteristic length greater than at least 130 $h$, which is much larger than the size of the very large scale motions expected in the channel flow. 
    The fact that this cutoff frequency should not overlap with range frequencies associated with the near-wall turbulence is also supported by its comparison to the expected range of flow frequencies within the channel flow, reporting a Strouhal number $Sr=\frac{fh}{U_\infty}=3\cdot10^{-3}$ (here $f=$ 0.9 Hz. 
    
%
    \item The temperature fluctuations expected in the thin foil are of a comparable order of magnitude to the IR camera noise, necessitating additional filtering. A 3D-Gaussian filter (in the image plane and in time) with a smoothing kernel has been applied to smooth the data. In both spatial directions, the size of the kernel is 2 ($\Delta x^+ \approx 6$), contributing to dampening fluctuations with the size of a single pixel, observed to be a major issue after the high-pass filter. The size of the kernel in the temporal dimension is set to 0.5 to improve the smoothing in the foil in-plane directions without increasing the radius of action of the kernel.
    \item Furthermore, a feature-based filtering, analogous to that done in Ref. \cite{raiola2017towards} can be applied. The temperature maps can be decomposed in POD modes and a low-order reconstruction with only the most energetic modes can be employed to remove incoherent features. The number of the modes to be retained has been identified with the elbow method \cite{cattell1966scree}. 
    \item To reduce field noise, due to minor non-uniformity in the gains the of IR camera sensing elements, an additional Gaussian filter with an elongated kernel was applied in the spanwise direction. This measure is taken to filter out residual striped patterns due to the acquisition mechanism of the IR camera, which scans the sensor in rows. 
\end{itemize}

\section{Results} \label{sec:results}

The experimental campaign to test this experimental configuration has been done by acquiring datasets of 12000 samples at 180 Hz, tuning the Joule effect power input to different levels to explore the balance between sufficiently strong thermal fluctuations and spurious effects due to natural convection. 
The current and voltage values employed in the different cases described in this work are collected in table \ref{tab:heatinglevels}.
Following the model from section \ref{sec.heattransfer} and the data filtering recipe of section \ref{sec:datafiltering}, the convective heat transfer coefficient $h_c$ has been quantified. 

\begin{table}[] 
\centering
\caption{\label{tab:heatinglevels}Input settings for the Joule current. }
\begin{indented}
\item[]\begin{tabular}{cccc}
\br
\multirow{2}{*}{$a_f$ ($\mu$m)} & Mean heating  & \multirow{2}{*}{Current (A)}  & \multirow{2}{*}{Voltage (V)}  \\
& level (K) &  &  \\ \mr
5 & 15 & 5.0 & 3.2\\
5 & 25 & 6.5 & 4.0\\
5 & 35 & 8 & 5.0\\
10 & 30 & 10 & 3.7\\
\br
\end{tabular}
\end{indented}
\end{table} 

\subsection{Results of the thermal model}

\begin{figure*}[ht]
\centering    
    \includegraphics[trim= 0 0.8cm 0 0.8cm, width=\linewidth]{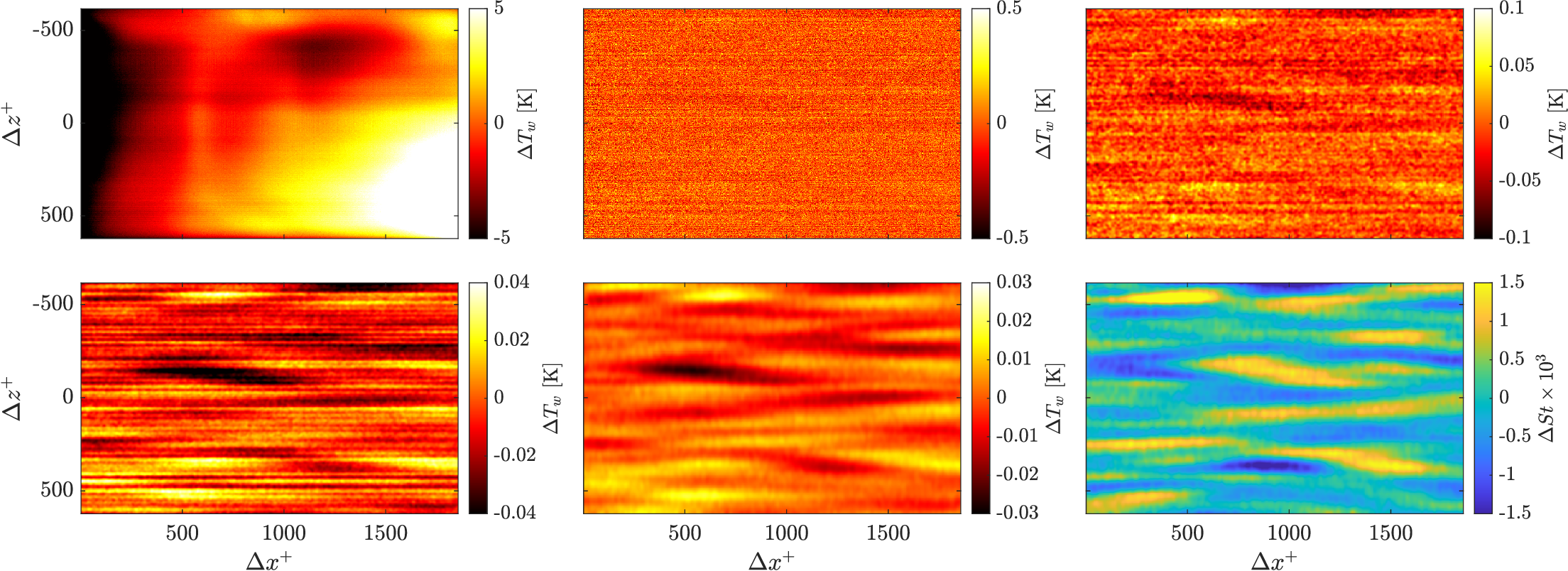}  
    \put(-490,170) {\small ($a$)}
    \put(-320,170) {\small ($b$)}
    \put(-165,170) {\small ($c$)}
    \put(-490,75) {\small ($d$)}
    \put(-320,75) {\small ($e$)}
    \put(-165,75) {\small ($f$)}
\caption{Maps of the thermal model at different filtering and processing stages in sequential order for an instantaneous field: ($a$) original, ($b$) high pass output, ($c$) 3D-Gaussian output, ($d$) POD output, ($e$) 3D-Gaussian output, ($f$) Stanton number map.}
\label{fig:Tmaps}
\end{figure*}

\begin{figure}[ht]
\centering    
    \includegraphics[trim= 0 0cm 0 0, width=0.8\linewidth]{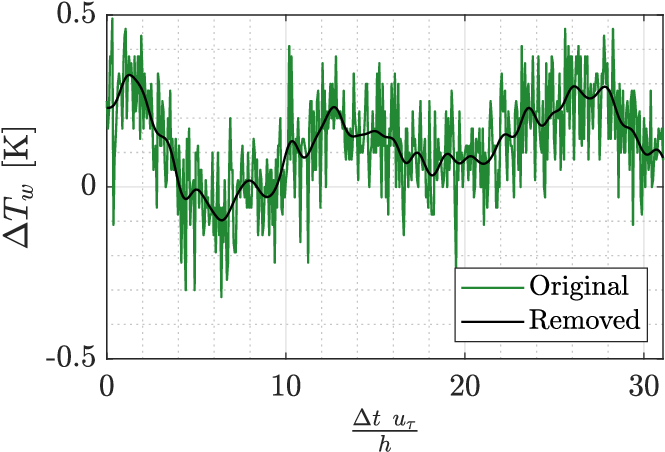}  
\caption{Original sequence of instantaneous temperature fluctuations of a pixel and removed low-frequency events.}
\label{fig:lowpassfiltering}
\end{figure}

The filtering sequence for an individual sample is depicted in figure \ref{fig:Tmaps}. These results correspond to measurements with a foil thickness of 5 $\mu$m. The input current is set to 8.0 A with a voltage supply of 5.0 V, resulting in a mean foil temperature of about 35 K above room temperature.
Although each filtering step may attenuate the temperature fluctuations, it is seen that the level of noise is progressively reduced. The high-pass filter has successfully removed the convective cells and other constant or low-frequency events not related to the turbulent features of the  channel-flow, as shown in figure \ref{fig:Tmaps} ($a$) and \ref{fig:lowpassfiltering}. However, the level of noise at this stage is still quite high as visible in figure \ref{fig:Tmaps} ($b$). Some streamwise patterns start being observed after the 3D-Gaussian filter in figure \ref{fig:Tmaps} ($c$). For further smoothing, the POD filter cuts the noisy tail of modes with about 50\% of the energy contained after the Gaussian filter according to the elbow criterion, see figure \ref{fig:Tmaps} ($d$). A Gaussian filter with actuation in the spanwise direction removes the remaining thin striped patterns, see fig. \ref{fig:Tmaps} ($e$). The resulting temperature map is used according to the energy balance in equation (\ref{eqn:energybalance}) to compute $h_c$, or the Stanton number $St=\frac{h_c}{\rho U_\infty c_p}$ in non-dimensional terms, employing the fluid density $\rho$ and specific heat capacity $c_p$. As seen in figure \ref{fig:Tmaps} ($f$), the outcome of this procedure results in instantaneous heat transfer maps with patterns that show a relation with the wall-bounded turbulence expected in the channel. These patterns are elongated and nearly aligned in the streamwise direction, with lengths in the range $\Delta x^+ = [500-1000]$ and a span of $\Delta z^+ = [50-100]$, with the superscript $^+$ indicating normalization with inner scales. These values are similar to those reported in Ref. \cite{del2003spectra}.

\subsection{Effect of heating}

\begin{figure}[ht]
\centering    
    \includegraphics[trim= 0 0cm 0 0, width=0.8\linewidth]{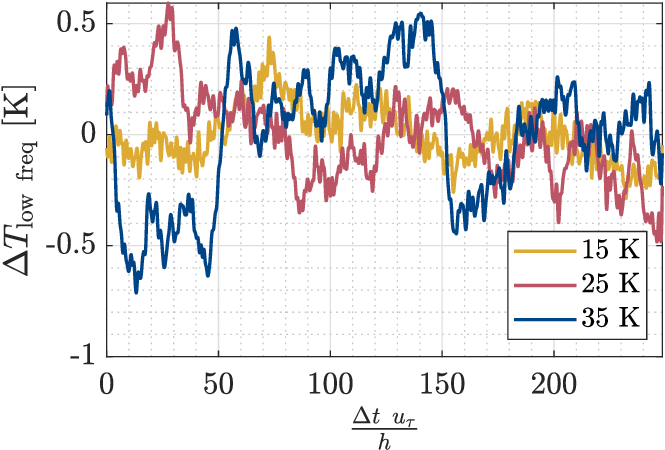}  
\caption{A sequence of low-frequency temperature fluctuations removed from a pixel with the high pass filter.}
\label{fig:lowpassfiltering_changeT}
\end{figure}

A sensitivity analysis of the power input effect is conducted, heating the foil to mean temperatures of about 15, 25 and 35 K above room temperature. Stronger heating leads to the development of more pronounced convective cells. As a result, the magnitude removed with the high pass filter increases with the power input, as shown in the sequences in figure \ref{fig:lowpassfiltering_changeT}.

\begin{figure*}[ht]
\centering    
    \includegraphics[trim= 0 0.8cm 0 0, width=\linewidth]{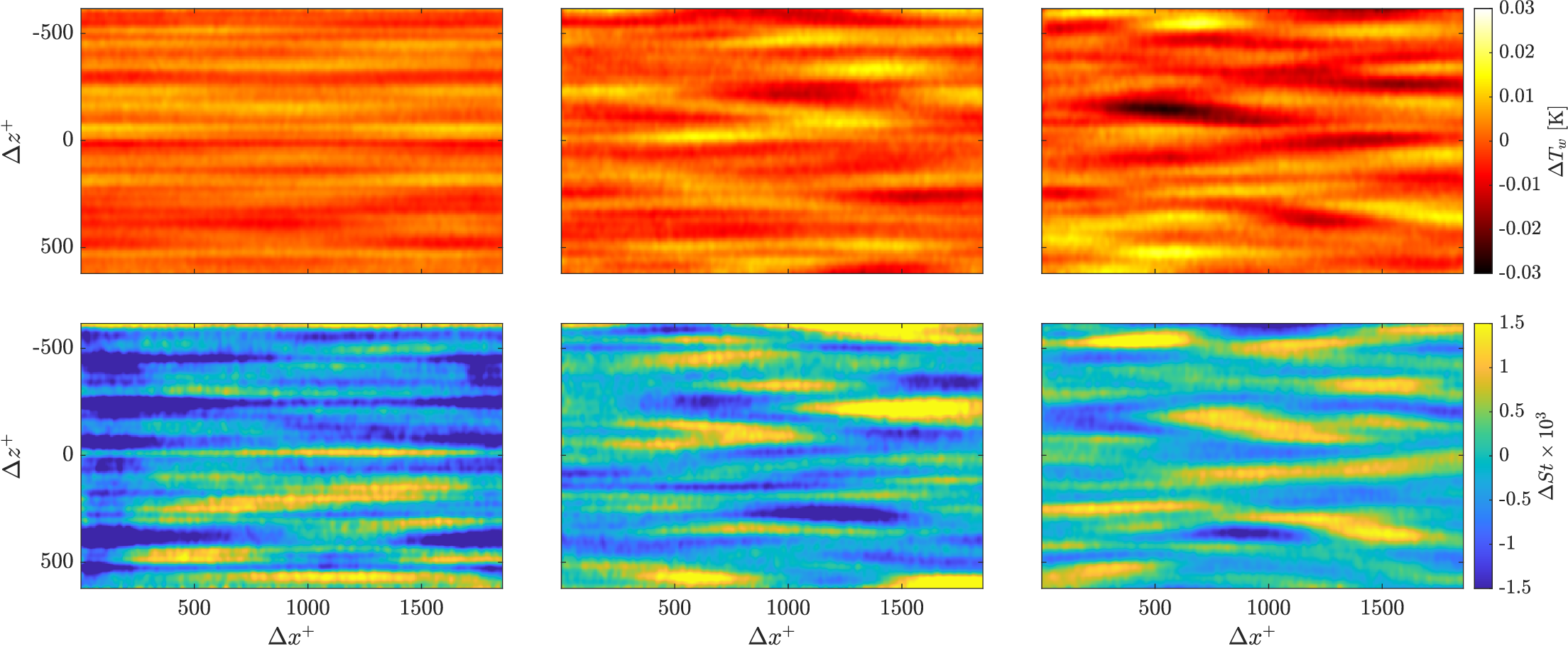}
    \put(-490,190) {\small ($a$)}
    \put(-327,190) {\small ($b$)}
    \put(-178,190) {\small ($c$)}
    \put(-490,90) {\small ($d$)}
    \put(-327,90) {\small ($e$)}
    \put(-178,90) {\small ($f$)} 
\caption{Instantaneous temperature map after the filtering process (top) and Stanton number map (bottom) at different heating conditions: from left to right ($a$) and ($d$) 3.2 V - 5.0 A (about 15 K), ($b$) and ($e$) 4.0 V - 6.5 A (about 25 K), ($c$) and ($f$) 5.0 V - 8.0 A (about 35 K, same as figure \ref{fig:Tmaps}$e-f$).}
\label{fig:dTcomparative}
\end{figure*}


Examples of filtered $T_{\rm w}$ and $St$ maps for these configurations are shown in figure \ref{fig:dTcomparative}. The temperature maps have been processed in the same manner for each power input level, with the POD filter threshold tailored to each case using the elbow method. Low heating levels make the temperature fluctuations less pronounced, while heating contributes to amplifying the signal. Once filtered, the temperature sequence for the lowest heating level (as in figure \ref{fig:dTcomparative} $a$) has a variance that is only 53\% of the one in the case with the highest heating level (figure \ref{fig:dTcomparative} $c$). The intermediate heating level retains 73\% of the variance corresponding to the highest heating level.

Beyond the different intensities of temperature fluctuations, $T_{\rm w}$ maps show differences in their patterns that strongly influence the $St$ patterns, as seen in the bottom row of figure \ref{fig:dTcomparative}. For the lowest heating level, the Stanton number patterns are still noticeably noisy and, from a qualitative inspection, seem less consistent with the physics of coherent structures in the near-wall region \cite{del2009estimation}. This sequence might necessitate further filtering, which may introduce more uncertainty. Further heating can magnify the temperature fluctuations, making the approach less susceptible to noise and facilitating obtaining patterns with physical meaning, as seen both in the $T_{\rm w}$ and $St$ maps. When the foil is heated 25 K above room temperature, the peaks are more evident than when it is heated 35 K. However, a higher level of heating shows less noisy Stanton number maps, with physical patterns being clearly identified. Quantitatively, the variance of these sequences can give a measure of the relative noise level among them. The intermediate and highest heating levels respectively retain 43\% and 25\% of the variance of the case with the lowest heating level. It must be considered though that natural convection is stronger in this latter case (although it is simple to filter with the strategy outlined in section \ref{sec:datafiltering}) and that the thermal expansion of the foil might produce undesired deformations. 

\subsection{Effect of foil thickness}

\begin{figure}[ht]
\centering    
    \includegraphics[trim= 0 0.8cm 0 0.8cm, width=0.8\linewidth]{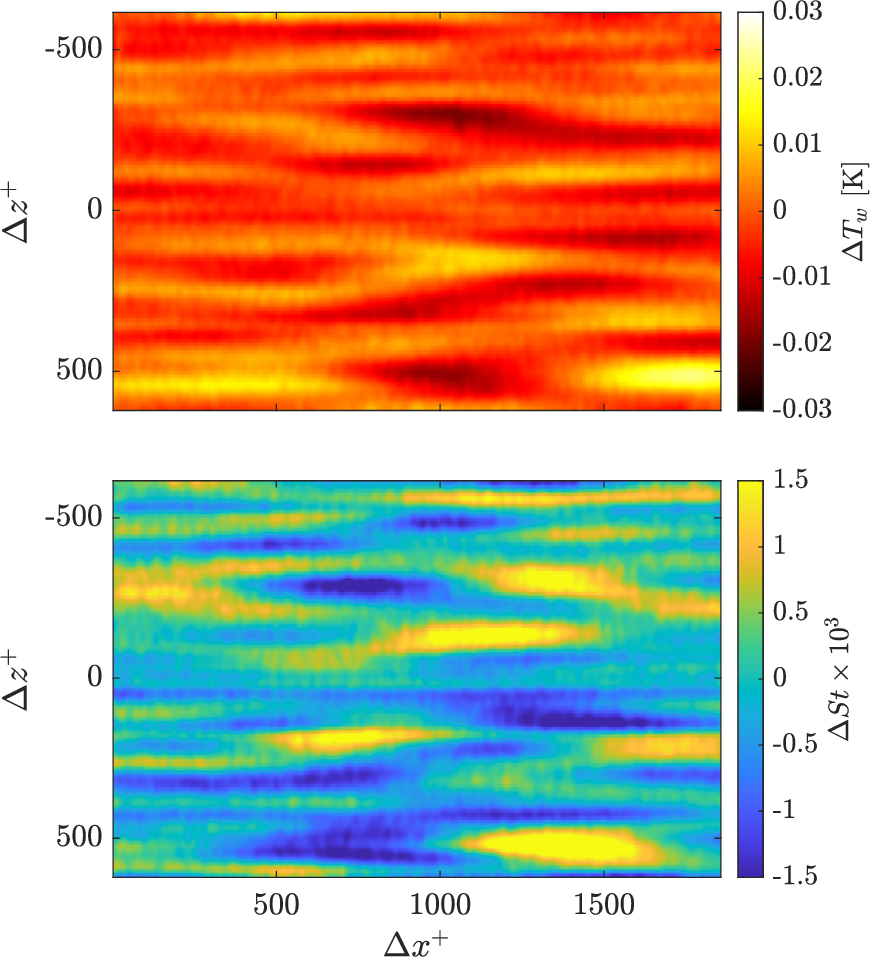}  
    \put(-190,195) {\small ($a$)}
    \put(-190,90) {\small ($b$)}
\caption{Instantaneous temperature map after the filtering process ($a$) and Stanton number map ($b$) from a 10 $\mu$m foil heated about 30 K with 3.7 V - 10.0 A.}
\label{fig:thinknesscomparative}
\end{figure}

We also conducted tests with a foil of thickness 10 $\mu$m. This comparison is performed using a similar heating to the case with higher heating (30 K) and the same filtering procedure and settings. Although the requirements of small Biot and large Fourier numbers are still met, this sensor may have less sensitivity due to the larger thermal inertia. A higher mass increases the thermal inertia and conduction, affecting the response time to temperature changes and distorting the temperature distribution on the surface. The map of filtered $T_{\rm w}$ and that of $St$ at a given time instant of the sequence are shown in figure \ref{fig:thinknesscomparative}, with the foil heated to a middle point between the cases in figures \ref{fig:dTcomparative} (centre and right). The temperature fluctuation map shows some patterns with shapes comparable to those with the 5 $\mu$m foil with the highest heating level that reasonably might represent near-wall turbulence. However, the magnitude of the peaks of these fluctuations is lower. More importantly, although not observed in a single map, a visual inspection of a temporal sequence shows slower temperature changes. This may affect the quantification of heat transfer, as temporal derivatives are directly involved. The $St$ map shows that the strongest fluctuations are well pronounced. However, other patterns, mainly representing noise rather than the wall-bounded turbulence physics in the near-wall region, remain stronger than those observed with the 5 $\mu$m foil. In light of these findings, one might expect that using a foil thinner than 5 $\mu$m would further reduce those effects.

\section{Uncertainty quantification} 
\label{sec:uncertainty}
The uncertainty of the model has been quantified with a Monte Carlo approach. The uncertainties indicated in table \ref{tab:foilparams} have been considered to define a 99\% confidence interval on a Gaussian distribution to generate multiple random combinations of these inputs. For the temperature maps and the ambient temperature, the NETD specified in the technical datasheet of the IR camera has been considered to be uniformly distributed through the POD modes and the Gaussian filters, thus the actual noise variance has been assumed as the NETD$^2$ times the percentage of POD modes retained and the pixelwise contribution of 
Gaussian filters. As such, the measurement uncertainty of fluctuating $St$ fields was quantified to be equal to 21.3\%, 13.8\%, 9.4\%, respectively for the cases $(a)$, $(b)$, and $(c)$, of figure \ref{fig:dTcomparative}. For the 10 $\mu$m foil, as reported in figure \ref{fig:thinknesscomparative}, it is equal to 11.6\%. Despite having the same uncertainties for the voltage, current and thickness, the higher quantities introduce less relative uncertainty than for the 5 $\mu$m foil while the larger thickness magnifies the uncertainty contribution of the temporal derivatives. These results support the choice of increased foil heating and smaller foil thickness as a means of improving measurement quality.

\section{Conclusions} \label{sec:conclusions}

The heat transfer fluctuations in an air channel flow are quantified through the energy balance of a heated thin foil embedded in the wall from time-resolved infrared measurements at high frequency. This paper provides an insight into the development of an experimental campaign to take this type of measurement in wall-bounded flows in a non-intrusive way. Temperature measurements are conducted employing IR thermography. The main challenges in tackling this experiment are the presence of high-frequency events to be captured, the noise of the experimental equipment and the thermal inertia and conduction of the foil. These heat transfer measurements have been made possible through the balance between different aspects and components of the problem, showing also examples of less convenient configurations. Some critical factors are the black matt coating, the power input and the foil thickness. Additionally, a filtering process is capable of isolating the characteristic turbulent phenomena of the channel from other effects and removing the high level of noise contained in the original temperature acquisitions. The impact of some parameters of the problem on this filtering process has been analysed. Summarising, the main guidelines obtained from this paper are the following:
\begin{itemize}
    \item It is recommended to remove spurious natural convection effects with high-pass filtering. It was shown that the natural convection effects could be suppressed for different heating levels, leading to similar results for all the conditions analysed in this note. 
    \item Measurement noise can be successfully removed with a feature-based POD filter coupled with a Gaussian filter to improve the data smoothness. 
    \item Foil thickness should be minimized whenever possible; measurement results are found to be weakly dependent on foil thickness provided the heating is strong enough to ensure sufficient signal-to-noise ratio. Note that when dealing with very thin foils, the presence of a thin layer of paint on the foil (necessary to enable precise IR measurements) must be taken into account when estimating the foil thermal inertia. 
\end{itemize}


\section*{Supplementary material}
Data and codes will be available upon publication.

\ack{}
A.C. acknowledges financial support from the Spanish Ministry of Universities under the FPU programme 2020.
This activity is part of the project ACCREDITATION (Grant No TED2021-131453B-I00), funded by MCIN/AEI/ 10.13039/501100011033 and by the “European Union NextGenerationEU/PRTR”.
This activity is part of the project EXCALIBUR (Grant No PID2022-138314NB-I00), funded by MCIU/AEI/ 10.13039/501100011033 and by “ERDF A way of making Europe”.
S.D. acknowledges funding from the European Research Council (ERC) under the European Union’s Horizon 2020 research and innovation programme (grant agreement no. 949085, NEXTFLOW).
G.C. acknowledges the MOST – Sustainable Mobility National Research Center and received funding from the European Union Next-GenerationEU (PIANO NAZIONALE DI RIPRESA E RESILIENZA (PNRR) – MISSIONE 4 COMPONENTE 2, INVESTIMENTO 1.4 – D.D. 1033 17/06/2022, CN00000023). Views and opinions expressed are, however, those of the authors only, and do not necessarily reflect those of the European Union or the ERC. Neither the European Union nor the granting authority can be held responsible for them.


\section*{References}
\begin{thebibliography}{9}






\bibitem{nakamura2009frequency}
Nakamura H., 2009, Frequency response and spatial resolution of a thin foil for heat transfer measurements using infrared thermography \emph{Int. J. Heat Mass Tran.}, \textbf{52}(21-22), 5040--5045.

\bibitem{nakamura2013quantitative}
Nakamura H. and Yamada S., 2013, Quantitative evaluation of spatio-temporal heat transfer to a turbulent air flow using a heated thin-foil \emph{Int. J. Heat Mass Tran.}, \textbf{64}, 892--902.

\bibitem{hetsroni1994heat}
Hetsroni G. and Rozenblit R., 1994, Heat transfer to a liquid—solid mixture in a flume \emph{Int J. Multiphas. Flow}, \textbf{20}, 671--689.

\bibitem{gurka2004detecting}
Gurka R., Liberzon A. and Hetsroni G., 2004, Detecting coherent patterns in a flume by using PIV and IR imaging techniques \emph{Exp. Fluids}, \textbf{37}, 230--236.

\bibitem{foroozan2023synchronized}
Foroozan F., G{\"u}emes A., Raiola M., Castellanos R., Discetti S. and Ianiro A, 2023, Synchronized measurement of instantaneous convective heat flux and velocity fields in wall-bounded flows \emph{Meas. Sci. Tech.}, \textbf{34}(12), 125301.

\bibitem{bergman2020fundamentals}
Bergman T. L., Lavine A. S., Incropera F. P. and DeWitt D. P.  2020 {\it Fundamentals of Heat and Mass Transfer}, John Wiley \& Sons

\bibitem{greco2014time}
Greco, C. S., Ianiro A. and Cardone G., 2014, Time and phase average heat transfer in single and twin circular synthetic impinging air jets \emph{Int. J. Heat Mass Tran.}, \textbf{73}, 776--788.

\bibitem{stafford2009characterizing}
Strafford J., Walsh E. and Egan V., 2009, Characterizing convective heat transfer using infrared thermography and the heated-thin-foil technique \emph{Meas. Sci. Tech.}, \textbf{20}(10), 105401.

\bibitem{torre2018evaluation}
Torre A.F.M., Ianiro A. Discetti S. and Carlomagno G.M., 2018, Evaluation of anisotropic tangential conduction in printed-circuit-board heated-thin-foil heat flux sensors \emph{Int. J. Heat Mass Tran.}, \textbf{127}, 1138--1146.

\bibitem{astarita2012infrared}
Astarita T. and Carlomagno G. M. 2013 {\it Infrared thermography for thermo-fluid-dynamics}, Springer, Berlin, Heidelberg.

\bibitem{raiola2017towards}
Raiola M., Greco C. S., Contino M., Discetti S. and Ianiro A., 2017, Towards enabling time-resolved measurements of turbulent convective heat transfer maps with IR thermography and a heated thin foil \emph{Int. J. Heat Mass Tran.}, \textbf{108}, 199--209.

\bibitem{gu2024denoising}
Gu F., Discetti S., Liu Y., Cao Z. and Peng D., 2024, Denoising image-based experimental data without clean targets based on deep autoencoders \emph{Exp. Therm. Fluid Sci.}, \textbf{156}, 111195.

\bibitem{serpieri2024conditioning}
Serpieri J., Cafiero G. and Iuso G., 2024, Conditioning Turbulent Channel Flows With Wall Plasma Jets \emph{AIAA aviation forum and ascend 2024}, p. 4384.

\bibitem{wong1977handbook}
Wong H. Y., 1977, \emph{Handbook of essential formulae and data on heat transfer for engineers}, Longman, London.

\bibitem{del2009estimation}
Del {\'A}lamo J. C. and Jim{\'e}nez, J., 2009, Estimation of turbulent convection velocities and corrections to Taylor's approximation \emph{J. Fluid Mech.}, \textbf{640}, 5--26.

\bibitem{raghu2006thermal}
Raghu O. and Philip J., 2006, Thermal properties of paint coatings on different backings using a scanning photo acoustic technique \emph{Meas. Sci. Tech.}, \textbf{17}(11), 2945.

\bibitem{susa2008influence}
Susa M., Maldaque X., Svaic S., Boras I. and Bendada A., 2008, The influence of surface coatings on the differences between numerical and experimental results for samples subject to a pulse thermography examination \emph{Proc. 9th Int. Conf. Quant. Infr. Thermogr. (QIRT)}, 491--497.

\bibitem{cattell1966scree}
Cattell R. B., 1966, The scree test for the number of factors \emph{Multivar. Behav. Res.}, \textbf{1}(2), 245--276.

\bibitem{del2003spectra}
Del {\'A}lamo, Juan C and Jim{\'e}nez, Javier, 2003, Spectra of the very large anisotropic scales in turbulent channels \emph{Phys. Fluids}, \textbf{15}(6), L41--L44.


\end{thebibliography}
\end{document}